\documentclass[fleqn,twoside]{article}
\usepackage{espcrc2}

\newcommand{\bei}{\begin{itemize}}
\newcommand{\eei}{\end{itemize}}
\newcommand{\beq}{\begin{equation}}
\newcommand{\eeq}{\end{equation}}
\newcommand{\beqn}{\begin{eqnarray}}
\newcommand{\eeqn}{\end{eqnarray}}
\newcommand{\beqns}{\begin{eqnarray*}}
\newcommand{\eeqns}{\end{eqnarray*}}

\newcommand{\intl}{\int\limits}

\newcommand{\mc}{\multicolumn}

\def\pc{$\%$}

\def\sf{spectral function}
\def\sfs{spectral functions}

\def\ee{$e^+e^-$}

\def\aqedZ{$\alpha(M_{Z}^2)$}

\def\daqedZ{$\Delta\alpha(M_{Z}^2)$}

\def\amuhadLO{$a_\mu^{\rm had,LO}$}
\def\tauto{$\tau^{-\!}\rightarrow\,$}
\def\nut{$\,\nu_\tau$}

\def\pipiz{$ \pi^-\pi^0 $}
\def\pitpiz{$ \pi^-3\pi^0 $}
\def\tpipiz{$ 2\pi^-\pi^+\pi^0 $}
\def\rar{\rightarrow}

\def\ie{{\it i.e.}} 
 
\def\via{via} 

\def\cf{{\em cf.}}
\def\rs{\raisebox{1.5ex}[-1.5ex]}

\usepackage{graphicx}
\usepackage[figuresright]{rotating}

\newcommand{\AmS}{{\protect\the\textfont2
  A\kern-.1667em\lower.5ex\hbox{M}\kern-.125emS}}

\title{Confronting Spectral Functions from \ee\ Annihilation 
	and $\tau$ Decays:
	Consequences for the Muon Magnetic Moment}

\author{A.~H\"ocker\address[LAL]{Laboratoire de l'Acc\'el\'erateur Lin\'eaire,\\
				IN2P3-CNRS et Universit\'e de Paris-Sud, 
				F-91898 Orsay, France \\
				{\footnotesize 
				(e-mail: hoecker@lal.in2p3.fr)}}%
        \thanks{Work done in collaboration with 
		M.~Davier, S.~Eidelman and Z.~Zhang},
}	
       
\begin{document}

\begin{abstract}
Vacuum polarization integrals involve the vector spectral functions which
can be experimentally determined from two sources:
({\it i}) \ee\ annihilation cross sections and ({\it ii}) hadronic $\tau$ 
decays. Recently results with comparable precision have become available 
from CMD-2 on one side, and ALEPH, CLEO and OPAL on the other. The comparison
of the respective spectral functions involves a correction from
isospin-breaking effects, which is evaluated. After the correction
it is found that the dominant $\pi\pi$ spectral functions do not 
agree within experimental and theoretical uncertainties. Some 
disagreement is also found for the $4\pi$ spectral functions. 
The consequences of these discrepancies for vacuum polarization
calculations are presented, with the emphasis on
the muon anomalous magnetic moment. Adding quadratically experimental
and theoretical uncertainties, we find that the full
Standard Model prediction of $a_\mu$ deviates from the recent 
BNL measurement at the level of 3.0 (\ee-based) and 0.9 ($\tau$-based) 
standard deviations.
\vspace{1pc}
\end{abstract}

\maketitle
\section{INTRODUCTION}

Hadronic vacuum polarization in the photon propagator plays an important 
role in the precision tests of the Standard Model. This is the case for the
evaluation of the electromagnetic coupling at the $Z$ mass scale,
$\alpha (M_Z^2)$, which receives a contribution 
$ \Delta\alpha_{\rm had}(M_{Z}^2)$ of the order of $2.8~10^{-2}$ 
that must be known to an accuracy of better than 1\% so that it does not
limit the accuracy on the indirect determination of the Higgs boson mass
from the measurement of $\sin ^2 \theta_W$. Another example is provided
by the anomalous magnetic moment $a_\mu=(g_\mu -2)/2$ of the muon, where the
hadronic vacuum polarization component is the leading contributor to the
uncertainty of the theoretical prediction.

Starting from Refs.~\cite{cabibbo,bouchiat} there is a long history of
calculating the contributions from hadronic vacuum polarization 
in these processes. As they cannot be obtained 
from first principles because of the low energy scale
involved, the computation relies on analyticity and unitarity so that the
relevant integrals can be expressed in terms of an experimentally determined
spectral function which is proportional to the cross section for \ee\
annihilation into hadrons. The accuracy of the calculations has therefore
followed the progress in the quality of the corresponding data~\cite{eidelman}.
Because the latter was not always suitable, it was deemed necessary to resort 
to other sources of information. One such possibility was the 
use~\cite{adh} of the vector spectral functions derived from the study 
of hadronic $\tau$ decays~\cite{aleph_vsf} for the energy range less 
than 1.8~GeV. Another one occurred when it was realized in the study of 
$\tau$ decays~\cite{aleph_asf} that perturbative QCD could be applied to 
energy scales as low as 1-2~GeV, thus offering a way to replace poor \ee\ 
data in some energy regions by a reliable and precise theoretical 
prescription~\cite{dh97,steinhauser,martin}. Finally, without any further 
theoretical assumption, it was proposed to use QCD sum rules~\cite{groote,dh98}
in order to improve the evaluation in energy regions dominated by
resonances where one has to rely on experimental data. 
Using these improvements the lowest-order hadronic contribution 
to $a_\mu$ was found to be~\cite{dh98}
\beq
    a_\mu^{\rm had,LO} \:=\: (692.4 \pm 6.2)~10^{-10}~.
\label{dh98}
\eeq
The complete theoretical prediction includes in addition QED, 
weak and higher order hadronic contributions.

The anomalous magnetic moment of the muon is experimentally
known to very high accuracy. Combined with the older, less precise results
from CERN~\cite{bailey}, the measurements from the E821 experiment at 
BNL~\cite{carey,brown,bnl},
including the most recent result~\cite{bnl_2002},
 yield 
\beq
    a_\mu^{\rm exp} \:=\: (11\,659\,203 \pm 8)~10^{-10}~,
\label{bnl}
\eeq
and are aiming at an ultimate precision of $4~10^{-10}$ in the future. 
The previous experimental result~\cite{bnl} was found to 
deviate from the theoretical prediction by 2.6~$\sigma$, 
but a large part of the discrepancy was 
originating from a sign mistake in the calculation of the small
contribution from the so-called light-by-light (LBL) scattering 
diagrams~\cite{kino_light,bij_light}. The new calculations of the
LBL contribution~\cite{knecht_light,kino_light_cor,bij_light_cor} have
reduced the discrepancy to a nonsignificant 1.6~$\sigma$ level. At any rate
it is clear that the presently achieved experimental accuracy already 
calls for a more precise evaluation of \amuhadLO.

New experimental and theoretical developments have prompted
the re-evaluation of the hadronic contributions presented in 
Ref.~\cite{dehz} and reported here:
\bei
\item 	new, precise results have been obtained at Novosibirsk with the CMD-2 
	detector in the region dominated by the $\rho$ resonance~\cite{cmd2},
	and more accurate R measurements have been 
	performed in Beijing with the BES detector in the 2-5 GeV energy 
	range~\cite{bes}.

\item 	new preliminary results are available from the final analysis of 
	$\tau$ decays with ALEPH using the full statistics accumulated at 
	LEP1~\cite{aleph_new}; also the information from the spectral 
	functions measured by CLEO~\cite{cleo_2pi,cleo_4pi} and 
	OPAL~\cite{opal} has been incorporated 
	in the analysis. 

\item 	new results on the evaluation of isospin breaking have been 
	produced~\cite{czyz,ecker1,ecker2}, thus providing a better 
	understanding of this critical area when relating vector $\tau$ 
	and isovector \ee\   spectral functions.
\eei

%
% ----------------  Muon Magnetic Anomaly -------------------
%
%
% --------- 
%
\section{MUON MAGNETIC ANOMALY}
\label{anomaly}

It is convenient to separate the Standard Model prediction for the
anomalous magnetic moment of the muon
into its different contributions,
\beq
    a_\mu^{\rm SM} \:=\: a_\mu^{\rm QED} + a_\mu^{\rm had} +
                             a_\mu^{\rm weak}~,
\eeq
with $
 a_\mu^{\rm had} \:=\: a_\mu^{\rm had,LO} + a_\mu^{\rm had,HO}
           + a_\mu^{\rm had,LBL}~,
$
where $a_\mu^{\rm QED}=(11\,658\,470.6\pm0.3)~10^{-10}$ is 
the pure electromagnetic contribution (see~\cite{hughes,cm} and references 
therein), \amuhadLO\ is the lowest-order contribution from hadronic 
vacuum polarization, $a_\mu^{\rm had,HO}=(-10.0\pm0.6)~10^{-10}$ 
is the corresponding higher-order part~\cite{krause2,adh}, 
and $a_\mu^{\rm weak}=(15.4\pm0.1\pm0.2)~10^{-10}$,
where the first error is the hadronic uncertainty and the second
is due to the Higgs mass range, accounts for corrections due to
exchange of the weakly interacting bosons up to two loops~\cite{amuweak}. 
For the LBL part we add the values for the pion-pole 
contribution~\cite{knecht_light,kino_light_cor,bij_light_cor} and the
other terms~\cite{kino_light_cor,bij_light_cor} to obtain
$a_\mu^{\rm had,LBL}=(8.6\pm3.5)~10^{-10}$.

By virtue of the analyticity of the 
vacuum polarization correlator, the contribution of the hadronic 
vacuum polarization to $a_\mu$ can be calculated \via\ the dispersion 
integral~\cite{rafael}
\beq\label{eq_int_amu}
    a_\mu^{\rm had,LO} \:=\: 
           \frac{\alpha^2(0)}{3\pi^2}
           \intl_{4m_\pi^2}^\infty\!\!ds\,\frac{K(s)}{s}R(s)~,
\eeq
where $K(s)$ is the QED kernel~\cite{rafael2} strongly emphasizing the 
low-energy spectral functions in the integral~(\ref{eq_int_amu}).
In effect, about 91\pc\ of the total contribution to \amuhadLO\ 
is accumulated at center-of-mass 
energies $\sqrt{s}$ below 1.8~GeV and 73\pc\ of \amuhadLO\ is covered by 
the two-pion final state which is dominated by the $\rho(770)$ 
resonance. 
In Eq.~(\ref{eq_int_amu}), $R(s)\equiv R^{(0)}(s)$ 
denotes the ratio of the 'bare' cross
section for \ee\ annihilation into hadrons to the pointlike muon-pair cross
section. At low mass-squared, $R(s)$ is taken from experiment.

%
% ------------------------- The ee Data at low energies -----------------
%
\section{DATA FROM \boldmath\ee\ ANNIHILATION}
\label{sec_dat_ee}

The exclusive low energy \ee\ cross sections have been measured mainly by 
experiments running at \ee\ colliders in Novosibirsk and Orsay. Due to the 
high hadron multiplicity at energies above $\sim 2.5$~GeV, the exclusive 
measurement of the respective hadronic final states is not practicable.
Consequently, the experiments at the high energy colliders have measured 
the total inclusive cross section ratio $R$. 

The most precise data from CMD-2 on the \ee$\rightarrow\pi^+\pi^-$
cross sections are now available in their final 
form~\cite{cmd2}. They differ from the preliminary ones, released 
two years ago~\cite{cmd2_prel}, mostly in the treatment of the radiative
corrections. The various changes resulted 
in a reduction of the cross section by about 1\% below the 
$\rho$ peak and 5\% above. The overall
systematic error of the final data is quoted to be 0.6\% and is
dominated by the uncertainties in the radiative corrections (0.4\%).
Agreement is observed between CMD-2 and the previous experiments
within the much larger uncertainties (2-10\%) quoted by the latter.

Large discrepancies are observed between the different data sets for 
the \ee$\rightarrow\pi^+\pi^-\pi^0\pi^0$ cross sections
(see Fig.~\ref{fig_2pi2pi0_eetau}).
These are probably related to problems in the calculation of the detection
efficiency, since
the efficiencies are small in general ($\sim 10-30$\%) and 
they are affected by uncertainties in the decay dynamics that is assumed 
in the Monte Carlo simulation. One could expect the more recent experiments 
(CMD-2~\cite{cmd2_2pi2pi0} and SND~\cite{snd_2pi2pi0}) to be more 
reliable in this context because of specific 
studies performed in order to identify the major decay processes involved.

For $e^+e^-\rightarrow\pi^+\pi^-\pi^+\pi^-$
the experiments agree reasonably well within
their quoted uncertainties (see Fig.~\ref{fig_4pi_eetau} in 
Section~\ref{sec_compsf}).

A detailed compilation and complete references of all the data used 
for this analysis are given in Ref.~\cite{dehz}.

%
% ------------------------- The tau Data -----------------
%
\section{DATA FROM HADRONIC \boldmath$\tau$ DECAYS}

Data from $\tau$ decays into two- and four-pion final states
\tauto\nut\pipiz, \tauto\nut\pitpiz\ and \tauto\nut\tpipiz,
are available from ALEPH~\cite{aleph_new,aleph_vsf}, 
CLEO~\cite{cleo_2pi,cleo_4pi} and OPAL~\cite{opal}.
The branching fraction $B_{\pi\pi^0}$ for the 
$\tau\rightarrow\nu_\tau\,\pi^-\pi^0~(\gamma)$ decay mode is of 
particular interest since it provides the normalization of the
corresponding spectral function. The new value~\cite{aleph_new},
$B_{\pi\pi^0}=(25.47 \pm 0.13)~\%$, turns out to be larger
than the previously published one~\cite{aleph_h} based on the 1991-93
LEP1 statistics, $(25.30 \pm 0.20)~\%$.

In the limit of isospin invariance, the corresponding \ee\ isovector
cross sections are calculated \via\ the isospin rotations
\begin{eqnarray}
\label{eq_cvc_2pi}
 \sigma_{e^+e^-\rightarrow\,\pi^+\pi^-}^{I=1}
        & = &
 \frac{4\pi\alpha^2}{s}\,v_{\pi^-\pi^0}~, \\[0.3cm]
\label{eq_cvc_4pi}
 \sigma_{e^+e^-\rightarrow\,\pi^+\pi^-\pi^+\pi^-}^{I=1} 
        & = &
             2\cdot\frac{4\pi\alpha^2}{s}\,
             v_{\pi^-\,3\pi^0}~, \\[0.3cm]
\label{eq_cvc_2pi2pi0}
 \sigma_{e^+e^-\rightarrow\,\pi^+\pi^-\pi^0\pi^0}^{I=1} 
        & = &
             \frac{4\pi\alpha^2}{s}\,
             \bigg[v_{2\pi^-\pi^+\pi^0} \\
	&&\hspace{1.1cm}
                  -\; v_{\pi^-\,3\pi^0}
             \bigg]~.
\end{eqnarray}
The $\tau$ \sf\ $v_V(s)$ for a given vector hadronic state $V$ is 
defined by~\cite{tsai}
\beqn
\label{eq_sf}
   v_V(s) 
   &\propto&          
              \frac{B(\tau^-\rightarrow \nu_\tau\,V^-)}
                   {B(\tau^-\rightarrow \nu_\tau\,e^-\,\bar{\nu}_e)}         
              \frac{d N_{V}}{N_{V}\,ds}\,\nonumber\\
   &&\hspace{0.0cm}\times\;	
              \left[ \left(1-\frac{s}{m_\tau^2}\right)^{\!\!2}\,
                     \left(1+\frac{2s}{m_\tau^2}\right) \right]^{-1},
\eeqn
where $|V_{ud}|=0.9748\pm0.0010$ (using~\cite{pdg2002}).
The \sfs\ are obtained from the corresponding invariant mass distributions,
after subtracting out the non-$\tau$ background and the feedthrough from 
other $\tau$ decay channels, and after a final unfolding from 
detector effects such as energy and angular resolutions, acceptance,
calibration and photon identification. Agreement within errors is 
observed for the three input spectral functions $v_{\pi^-\pi^0}$
(ALEPH, CLEO, OPAL), so that we use their weighted average in 
the following.

%
% ------------------------- ee Radiative corrections ---------------------
%
\section{RADIATIVE CORRECTIONS FOR \boldmath\ee\ DATA}
\label{sec_rad}

Radiative corrections applied to the measured \ee\ cross sections are an
important step in the experimental analyses. They involve the consideration
of several physical processes and lead to large corrections. We stress
that the evaluation of the integral in Eq.~(\ref{eq_int_amu}) requires
the use of the 'bare' hadronic cross section, so that the input data must
be analyzed with care in this respect.
Several steps are to be considered:
\bei
\item 	Corrections are applied to the luminosity determination, based
	on large-angle Bhabha scattering and muon-pair production in 
	the low-energy experiments, and small-angle Bhabha scattering at 
	high energies. These processes are usually corrected for external
	radiation, vertex corrections and vacuum polarization from lepton 
	loops.

\item 	The hadronic cross sections given by the experiments are in general
	corrected for initial state radiation and the effect of loops 
	at the electron vertex.

\item 	The vacuum polarization correction in the photon propagator is
	a more delicate point. The cross sections need to be fully corrected 
	for our use, {\it i.e.} $\sigma_{\rm bare}= \sigma_{\rm dressed}
          (\alpha(0)/\alpha(s))^{2}$,
	where $\sigma_{\rm dressed}$ is the measured cross section 
	already corrected for initial state radiation, 
	and $\alpha(s)$ is obtained from resummation of the lowest-order
	evaluation, giving $\alpha(s)= \alpha(0)/
        (1 -\Delta\alpha_{\rm lep}(s) -\Delta\alpha_{\rm had}(s))$.
	Whereas $\Delta\alpha_{\rm lep}(s)$ can be calculated analytically,
	$\Delta\alpha_{\rm had}(s)$ is related by analyticity and unitarity 
	to a dispersion integral, akin to Eq.~(\ref{eq_int_amu}). Since the 
	hadronic correction involves the knowledge of $R(s)$ at 
	all energies, including those where the measurements are made, 
	the procedure has to be iterative, and requires experimental as 
	well as theoretical information over a large energy range.

	The new data from CMD-2~\cite{cmd2} are explicitly
	corrected for both leptonic and hadronic vacuum polarization effects,
	whereas the preliminary data from the same experiment~\cite{cmd2_prel} 
	and data from other experiments 
	were not (see follow up of this discussion in Ref.~\cite{dehz})

\item 	In Eq.~(\ref{eq_int_amu}) one must incorporate in $R(s)$ the
	contributions of all hadronic states produced at the energy 
	$\sqrt{s}$.
	In particular, radiative effects in the hadronic final state
	must be considered, \ie, final states such as $V+\gamma$ (FSR) have 
	to be included. While FSR has been added to 
	the newest CMD-2 data~\cite{cmd2}, this is not the case for 
	the older data and thus has to be corrected for~\cite{dehz}
	``by hand'', using an analytical expression computed in
	scalar QED (point-like pions)~\cite{jeger_rad}. 

\eei

In summary, we correct each \ee\ experimental result, but those from
CMD-2 ($\pi\pi$), by the factor $C_{\rm HVP}\cdot C_{\rm FSR}$, 
where $C_{\rm HVP}$
accounts for hadronic vaccum polarization and $C_{\rm FSR}$ stands
for the FSR correction.
We assign uncertainties of $50\%$ (vacuum polarization) and $100\%$
(FSR corrections), which are considered to be 
fully correlated between all channels to which the corrections apply.

%
% ------------------------------ CVC ---------------------
%
\section{ISOSPIN BREAKING IN \boldmath\ee\ AND \boldmath$\tau$ 
	 SPECTRAL FUNCTIONS}
\label{sec_cvc}

\begin{table}[t]
\setlength{\tabcolsep}{-0.08pc}
\caption{Expected sources of isospin symmetry breaking between 
         \ee\ and $\tau$ \sfs\ in the $2\pi$ and $4\pi$ channels, 
         and the corresponding corrections
         to \amuhadLO\ as obtained from $\tau$ data. }
\label{tab_isobreak}
\begin{tabular}{@{}lc} \hline
&\\[-0.3cm]
Sources of Isospin 	& \mc{1}{c}{$\Delta$\amuhadLO~($10^{-10}$)}      \\
Symmetry Breaking 
		& $\pi^+\pi^-$\\[0.15cm]
\hline
&\\[-0.3cm]
Short distance rad. corr.
	& $-12.1\pm0.3 $       \\
Long distance rad. corr.
	& $-1.0 $   \\
$m_{\pi^-}\ne m_{\pi^0}$ ($\beta$ in cross section)
	& $-7.0$           \\ 
$m_{\pi^-}\ne m_{\pi^0}$ ($\beta$ in $\rho$ width)    
	& $+4.2$            \\ 
$m_{\rho^-}\ne m_{\rho^0}$               
	& $0 \pm 2.0$        \\
$\rho-\omega$ interference              
	& $+3.5 \pm 0.6$     \\
Electromagnetic decay modes                   
	& $-1.4 \pm 1.2$\\[0.15cm]
\hline 
&\\[-0.3cm]
Sum                                     
	& $-13.8 \pm 2.4$  \\[0.15cm]
\hline
\end{tabular}
\end{table} 
The relationships~(\ref{eq_cvc_2pi}), (\ref{eq_cvc_4pi}) and 
(\ref{eq_cvc_2pi2pi0}) between \ee\ and $\tau$ spectral functions only hold
in the limit of exact isospin invariance. This is the Conserved Vector 
Current (CVC) property of weak decays. It follows from the factorization
of strong interaction physics as produced through the $\gamma$ and $W$
propagators out of the QCD vacuum.
However, we know that we must expect symmetry 
breaking at some level from electromagnetic effects and even in QCD  
because of the up and down quark mass splitting. Since the normalization
of the $\tau$ spectral functions is experimentally known at the 0.5\%
level, it is clear that isospin-breaking effects must be carefully examined
if one wants this precision to be maintained in the vacuum polarization 
integrals. 

Because of the dominance of the $\pi\pi$ contribution in the energy range 
of interest for $\tau$ data, we discuss mainly this channel,
following our earlier analysis~\cite{adh}. The corrections on \amuhadLO\
from isospin breaking are given in Table~\ref{tab_isobreak}. 
A more complete discussion, in particular with respect to the 
corrections previously applied is given in Ref.~\cite{dehz}.

The dominant contribution to the electroweak radiative corrections 
stems from the short distance correction to the effective 
four-fermion coupling $\tau^-\rightarrow\nu_\tau(d\bar{u})^-$
enhancing the $\tau$ amplitude by the factor 
$S^{\rm had}_{\rm{EW}}=1.0194$~\cite{marciano-sirlin}.
This correction leaves out the possibility of sizeable contributions 
from virtual loops. This problem was studied 
in Ref.~\cite{ecker1} within a model based on Chiral Perturbation Theory.
In this way the correct low-energy hadronic structure is implemented and
a consistent framework has been set up to calculate electroweak and strong
processes, such as the radiative corrections in the 
$\tau\rightarrow\nu_\tau\pi^-\pi^0$ decay. Their new analysis~\cite{ecker2} 
directly applies to the inclusive radiative rate,
$\tau\rightarrow\nu_\tau\,\pi^-\pi^0~(\gamma)$, as measured by the
experiments. The relation between the Born level \ee\ \sf\   and 
the $\tau$ \sf\   reads~\cite{ecker2} 
\beq
\label{eq:cvc_break}
v_{\pi^+\pi^-} \, = \, \frac{1}{G_{\rm EM}}
		\,\frac{\beta_0^3}{\beta_-^3}
               \,\left|\frac{F_\pi^0}{F_\pi^-}
		\right|^2 v_{\pi^-\pi^0(\gamma)}~,
\eeq
where $G_{\rm EM}\equiv G_{\rm EM}(s)$ is the long-distance radiative  
correction involving both real photon emission and virtual loops 
(the infrared divergence cancels in the sum). Note that the 
short-distance $S_{\rm EW}$ correction, discussed above,
is already applied in the definition of $v_-$ 
({\it cf.} Eq.~(\ref{eq_sf})), but its value differs from 
the one given in Ref.~\cite{marciano-sirlin},
because subleading quark-level and hadron-level contributions should not
be added, as double counting would occur. The correct expression for the
$\pi^- \pi^0$ mode therefore reads
\beq
\label{sew_rho}
  \frac{S^{\rm had}_{\rm{EW}}G_{\rm EM}}
           {S^{\rm sub,lep}_{\rm EW}}
                    =(1.0233\pm0.0006)\cdot G_{\rm EM}~,
\eeq
the subleading hadronic corrections being now incorporated in the
mass-squared-dependent $G_{\rm EM}$ factor.
Equation~(\ref{eq:cvc_break}) explicitly corrects for the mass 
difference between neutral and charged pions affecting the cross section
and the width of the $\rho$.

The different contributions to the isospin-breaking corrections
are shown in the second column of Table~\ref{tab_isobreak}.
The dominant uncertainty stems from the $\rho^\pm$-$\rho^0$
mass difference. 

Since the integral~(\ref{eq_int_amu}) requires as input the \ee\ \sf\ 
including FSR photon emission, a final correction is necessary. 
It is identical to that applied in the CMD-2 analysis~\cite{cmd2,jeger_rad}
(\cf\   Section~\ref{sec_rad}). The total correction to the $\tau$ 
result amounts to $(-9.3 \pm 2.4)~10^{-10}$.

There exists no comparable study of isospin breaking in the $4\pi$
channels. Only kinematic corrections resulting from the pion mass difference
have been considered so far~\cite{czyz}, which we have applied in
this analysis. It creates shifts of $-0.7~10^{-10}$ ($-3.8\%$) and
$+0.1~10^{-10}$ ($+1.1\%$) for $2\pi^+2\pi^-$ and $\pi^+\pi^-2\pi^0$,
respectively. 

%
% ------------------------------ CVC tests ---------------------
%
\section{COMPARISON OF \boldmath\ee\ AND \boldmath$\tau$ SPECTRAL FUNCTIONS}

The \ee\ and the isospin-breaking corrected $\tau$ \sfs\ can be 
directly compared for the dominant $2\pi$ and $4\pi$ final states. For
the $2\pi$ channel, the $\rho$-dominated form factor falls off very
rapidly at high energy so that the comparison can be performed 
in practice over the full energy range of interest. The situation
is different for the $4\pi$ channels where the $\tau$ decay kinematics
limits the exercise to energies less than $\sim$~1.6~GeV, with only
limited statistics beyond.

\label{sec_compsf}

\begin{figure}[t]
\includegraphics[width=17.5pc]{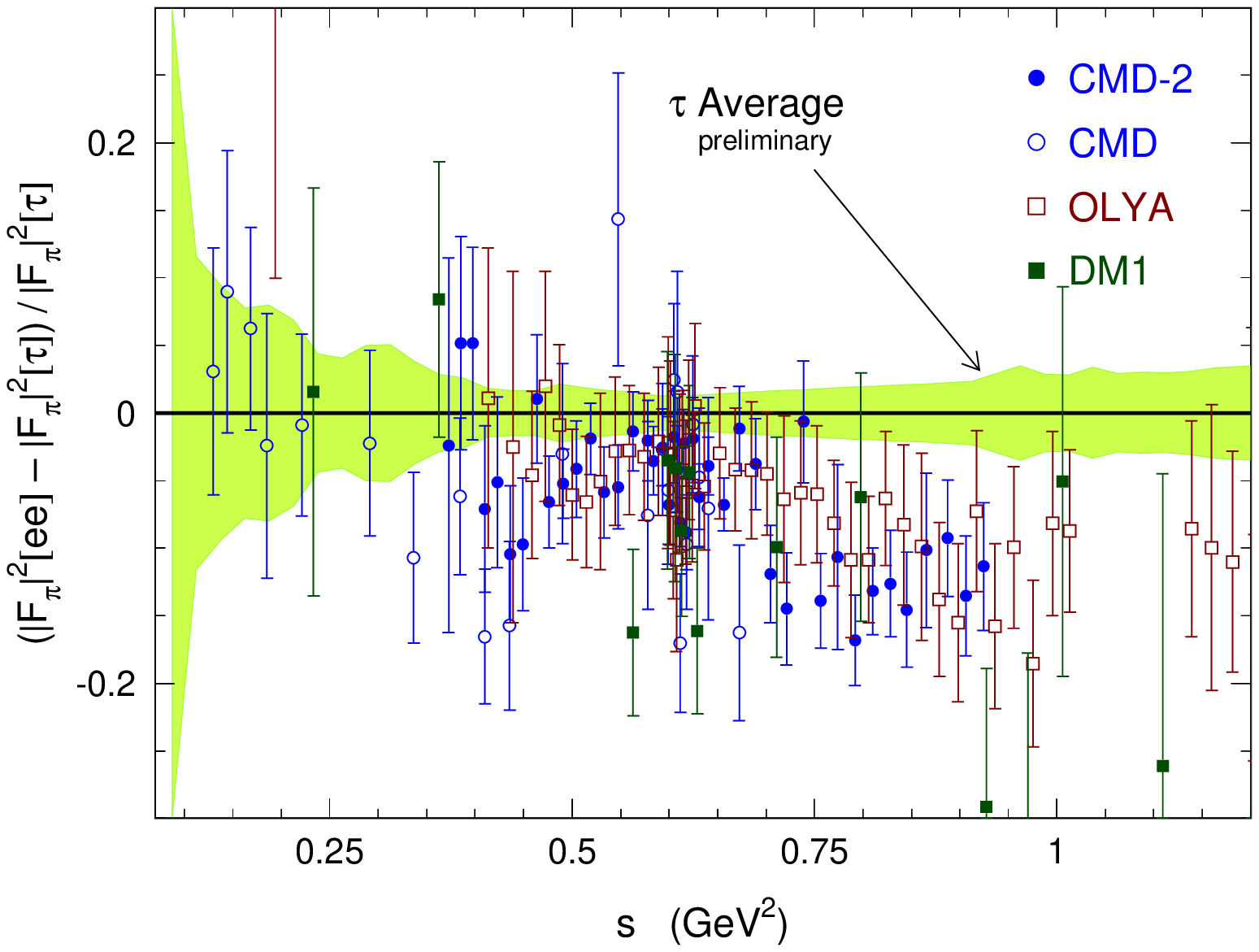}
\vspace{-0.7cm}
\caption[.]{Relative comparison of the $\pi^+\pi^-$ \sfs\
    	from \ee\  and isospin-breaking corrected $\tau$ data, 
	expressed as a ratio to the $\tau$ \sf.
	The band shows the uncertainty on the latter.}
\label{fig_2pi_comp}
\end{figure}
\begin{figure}[t]
\includegraphics[width=17.5pc]{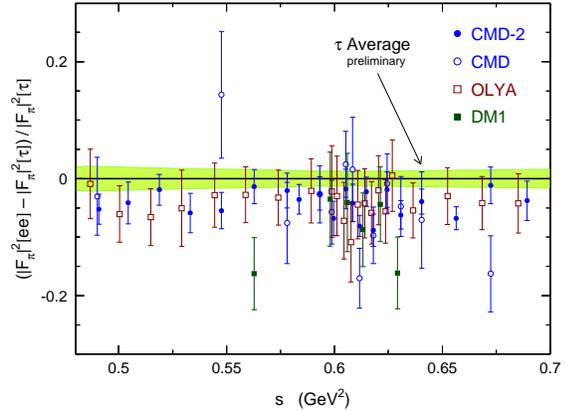}
\vspace{-0.7cm}
\caption[.]{Relative comparison in the $\rho$ region of the 
	$\pi^+\pi^-$ \sfs\
    	from \ee\  and isospin-breaking corrected $\tau$ data, 
	expressed as a ratio to the
     	$\tau$ \sf. The band shows the uncertainty on the latter.}
\label{fig_2pi_comp_zoom}
\end{figure}
\begin{figure*}[p]
\begin{center}
\includegraphics[width=34pc]{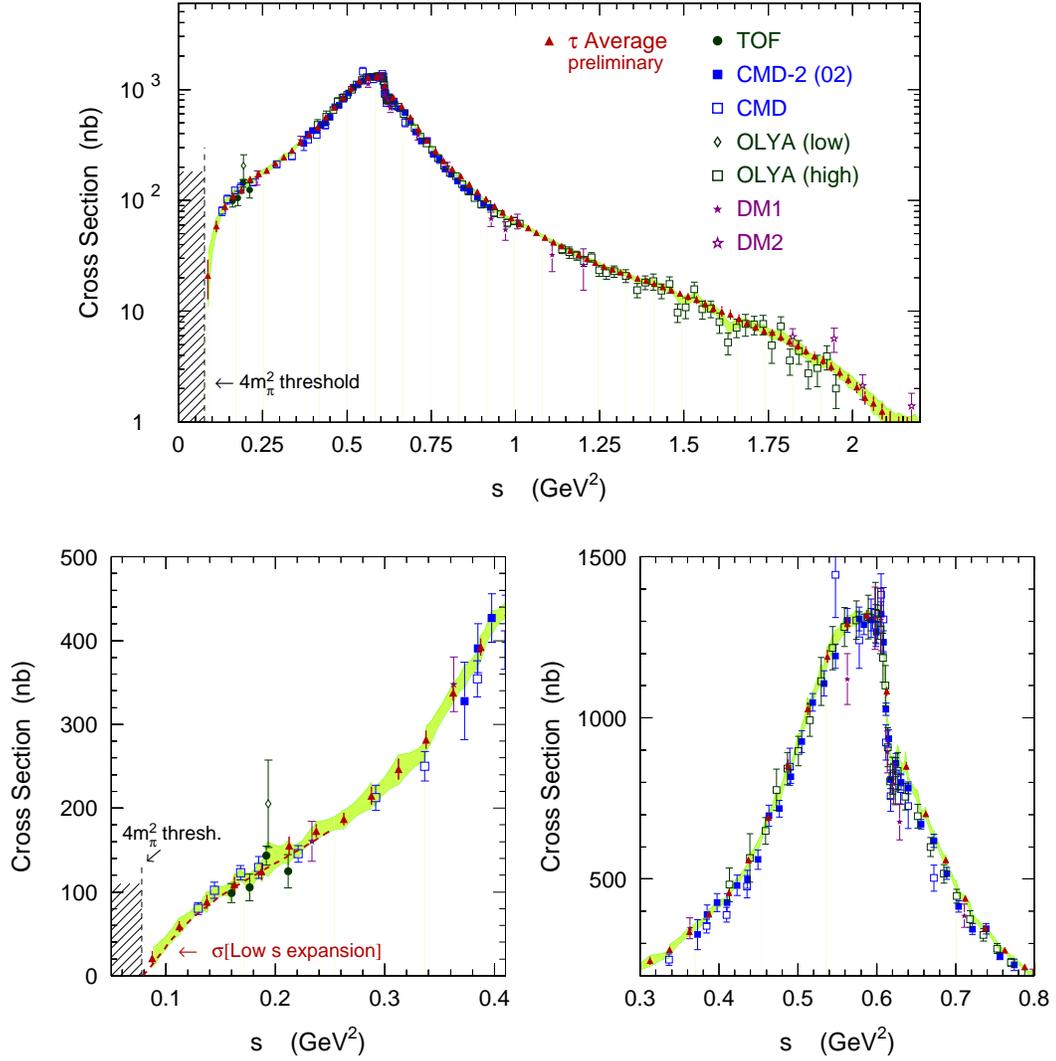}
\vspace{-0.7cm}
\end{center}
\caption[.]{Comparison of the $\pi^+\pi^-$ \sfs\
    	from \ee\  and isospin-breaking corrected $\tau$ data, 
	expressed as \ee\ cross sections. The band indicates the 
	combined \ee\ and $\tau$ result within $1\sigma$ errors.
	It is given for illustration purpose only. }
\label{fig_2pi_eetau}
\end{figure*}
\begin{figure}[t]
\includegraphics[width=17.5pc]{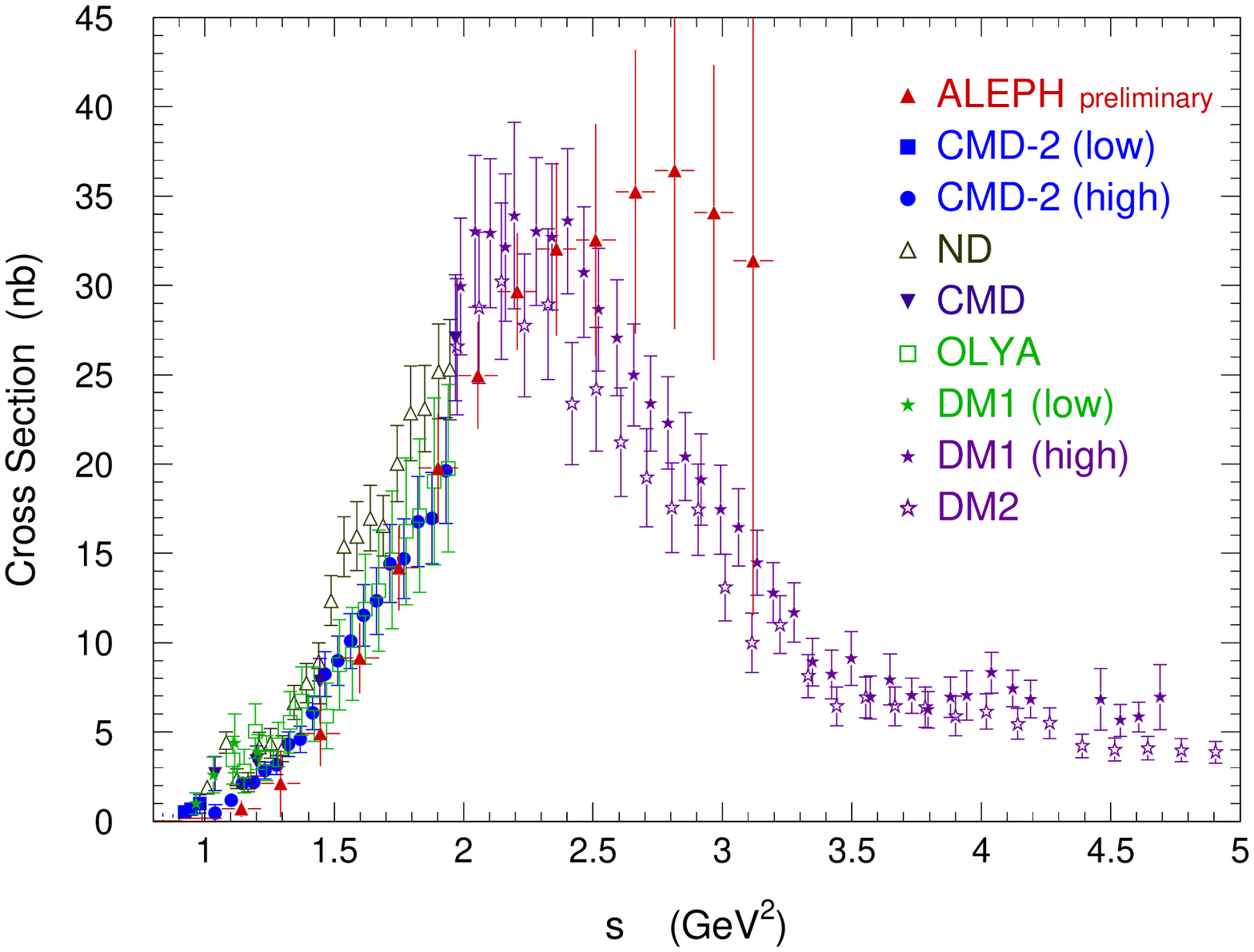}
\vspace{-0.7cm}
\caption[.]{Comparison of the $2\pi^+2\pi^-$ \sfs\
    	from \ee\ and isospin-breaking corrected $\tau$ data,
	expressed as \ee\ cross sections.}
\label{fig_4pi_eetau}
\end{figure}
\begin{figure}[t]
\includegraphics[width=17.5pc]{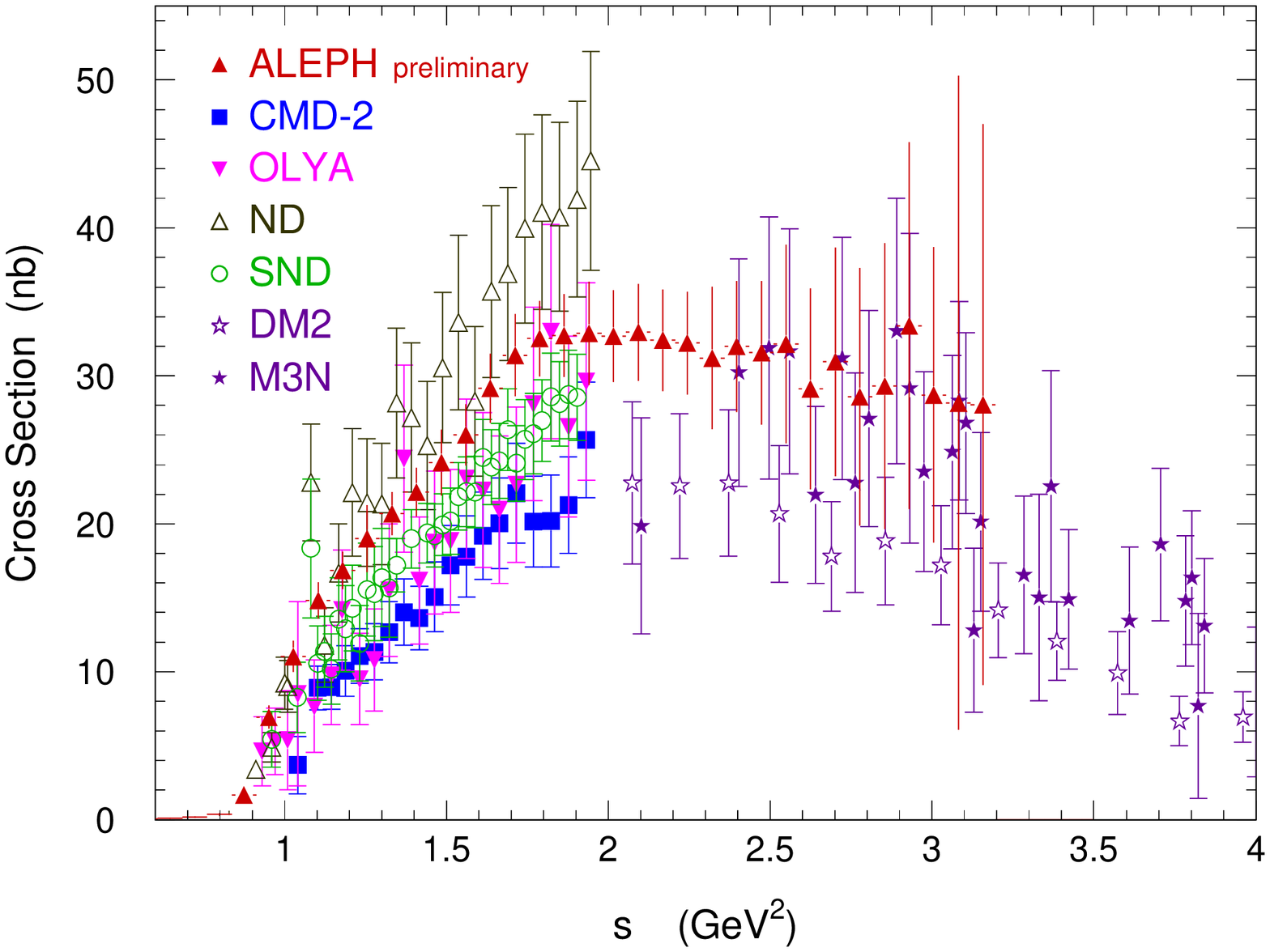}
\vspace{-0.7cm}
\caption[.]{Comparison of the $\pi^+\pi^-2\pi^0$ \sfs\
    	from \ee\ and isospin-breaking corrected $\tau$ data,
	expressed as \ee\ cross sections.}
\label{fig_2pi2pi0_eetau}
\end{figure}
Fig.~\ref{fig_2pi_eetau} shows the comparison for the $2\pi$ \sfs. 
Visually, the agreement seems satisfactory, however the large 
dynamical range involved
does not permit an accurate test. To do so, the \ee\ data are plotted
as a point-by-point ratio to the $\tau$ \sf\ in Fig.~\ref{fig_2pi_comp}, 
and enlarged in Fig.~\ref{fig_2pi_comp_zoom}, to better emphasize the 
region of the $\rho$ peak. The \ee\ data are significantly lower by 2-3\% 
below the peak, the discrepancy increasing to about 10\% in the 
0.9-1.0~GeV region.

The comparison for the $4\pi$ cross sections is given 
in Fig.~\ref{fig_4pi_eetau}
for the $2\pi^+2\pi^-$ channel and in Fig.~\ref{fig_2pi2pi0_eetau} for 
$\pi^+\pi^-2\pi^0$. The latter suffers from
large differences between the results from the different \ee\ 
experiments. The $\tau$ data, combining two measured \sfs\ according
to Eq.~(\ref{eq_cvc_2pi2pi0}) and corrected for isospin breaking as
discussed in Section~\ref{sec_cvc}, lie somewhat in between with large
uncertainties above 1.4~GeV because of the lack of statistics and a large 
feedthrough background in the $\tau\rightarrow\nu_\tau\,\pi^-3\pi^0$
mode. In spite of these difficulties the  $\pi^-3\pi^0$ \sf\ is in agreement
with \ee\ data as can be seen in Fig.~\ref{fig_4pi_eetau}. It is clear
that intrinsic discrepancies exist among the \ee\ experiments and that a
quantitative test of CVC in the $\pi^+\pi^-2\pi^0$ channel is
premature.

\subsection{Branching Ratios in $\tau$ Decays and CVC}
\label{sec_brcvc}

A convenient way to assess the compatibility between \ee\ and $\tau$
\sfs\ proceeds with the evaluation of $\tau$ decay fractions using
the relevant \ee\ \sfs\ as input. All the isospin-breaking corrections 
discussed in Section~\ref{sec_cvc} are included. The advantage of this
procedure is to allow a quantitative comparison using a single number.
The weighting of the \sf\ is however different from the vacuum
polarization kernels. Using the branching fraction
$B(\tau^-\rightarrow \nu_\tau\,e^-\,\bar{\nu}_e)\,=\,(17.810 \pm 0.039)\%$,
obtained assuming leptonic universality in the charged weak 
current~\cite{aleph_new}, the results for the main channels are given
in Table~\ref{tab_brcvc}. The errors quoted for the CVC values are split
into uncertainties from ({\it i}) the experimental input
(the \ee\ annihilation cross sections) and the numerical integration procedure,
({\it ii}) the missing radiative corrections applied to the relevant \ee\ data,
and ({\it iii}) the isospin-breaking corrections when relating $\tau$ and \ee\
\sfs. The values for the $\tau$ branching ratios involve 
measurements~\cite{aleph_new,cleo_bpipi0,opal_bpipi0}
given without charged hadron identification, {\it i.e.}, for the
$h\pi^0\nu_\tau$, $h3\pi^0\nu_\tau$ and $3h\pi^0\nu_\tau$ final states. 
The corresponding channels with charged kaons have been 
measured~\cite{aleph_ksum,cleo_kpi0} and their
contributions can be subtracted out in order to obtain the pure pionic
modes. 
\begin{table*}[t]
\caption{Branching fractions of $\tau$ vector decays into 	
	2 and 4 pions in the final state. Second column: world 
	average. Third column: inferred from \ee\ spectral
	functions using the isospin 
	relations~(\ref{eq_cvc_2pi}-\ref{eq_cvc_2pi2pi0})
        and correcting for isospin breaking. The experimental error
	of the $\pi^+\pi^-$ CVC value contains an absolute 
	procedural integration error of $\,0.08\%$.
	Experimental errors, including uncertainties on the integration 
        procedure, and theoretical (missing radiative corrections for \ee,
	and isospin-breaking corrections and $V_{ud}$ for $\tau$)
	are shown separately.
	Right column: differences between the
	direct measurements in $\tau$ decays and the CVC evaluations,
        where the separate errors have been added in quadrature.}
\label{tab_brcvc}
\setlength{\tabcolsep}{1.08pc}
{\normalsize
\begin{tabular}{lrrr} \hline 
&&& \\[-0.3cm]
		& \mc{3}{c}{Branching fractions  (in \%)} \\
\rs{~~~~~Mode} 	& \mc{1}{c}{$\tau$ data} 	
		& \mc{1}{c}{$e^+e^-$ via CVC} & $\Delta(\tau-e^+e^-)$ 
\\[0.15cm]
\hline
&&& \\[-0.3cm]
\mc{1}{l}{$~\tau^-\rar\nu_\tau\pi^-\pi^0$}
		& $25.46 \pm 0.12$ 
		& $23.98 \pm \underbrace{0.25_{\rm exp}	
			\pm 0.11_{\rm rad}\pm 0.12_{\rm SU(2)}}_{0.30}$ 
		& $+1.48 \pm 0.32$ 
	\\[0.7cm]
\mc{1}{l}{$~\tau^-\rar\nu_\tau\pi^-3\pi^0$}
		& $ 1.01 \pm 0.08$ 
		& $ 1.09 \pm \underbrace{0.06_{\rm exp}
		        \pm 0.02_{\rm rad}\pm 0.05_{\rm SU(2)}}_{0.08}$ 
		& $-0.08 \pm 0.11$ 
	\\[0.7cm]
\mc{1}{l}{$~\tau^-\rar\nu_\tau2\pi^-\pi^+\pi^0$}
		& $ 4.54 \pm 0.13$ 
		& $ 3.63 \pm \underbrace{0.19_{\rm exp}
			\pm 0.04_{\rm rad}\pm 0.09_{\rm SU(2)}}_{0.21}$ 
		& $+0.91 \pm 0.25$ 
	\\[0.7cm]
 \hline
\end{tabular}
}
\end{table*} 
As expected from the preceding discussion, a large discrepancy is 
observed for the $\tau\rightarrow \nu_\tau\,\pi^-\pi^0$ 
branching ratio, with a difference of 
$(-1.48\pm0.12_\tau\pm0.25_{\rm ee}\pm0.11_{\rm rad}
\pm0.12_{\rm SU(2)})\%$, 
where the uncertainties are from the $\tau$ branching ratio, 
\ee\ cross sections, \ee\ missing radiative corrections and isospin-breaking 
corrections (including the uncertainty on $V_{ud}$), respectively. 
Adding all errors in quadrature, the effect represents a 4.6~$\sigma$ 
discrepancy. Since the disagreement between \ee\ and $\tau$ \sfs\ is 
more pronounced at energies above 750~MeV, we expect a smaller discrepancy
in the calculation of \amuhadLO\ because of the steeply falling kernel
$K(s)$ in this case. 

The situation in the $4\pi$ channels is different. Agreement is
observed for the $\pi^-3\pi^0$ mode within an accuracy of $11\%$, however
the comparison is not satisfactory for the $2\pi^-\pi^+\pi^0$ mode. 
In the latter case, the relative difference is very large, 
$(22\pm6)$\%, compared to any reasonable level of isospin symmetry 
breaking. As such, it rather points to experimental problems that have
to be investigated.

%
% -----------------Specific analytical contributions ---------------------
%

\section{SPECIFIC CONTRIBUTIONS}

In some energy regions where data information is scarce and reliable 
theoretical predictions are available, we use analytical contributions
to extend the experimental integral. Also, the treatment of narrow
resonances involves a specific procedure.
%
% ---------------
%
\subsection{The $\pi^+\pi^-$ Threshold Region}

\begin{figure}[t]
\includegraphics[width=17.5pc]{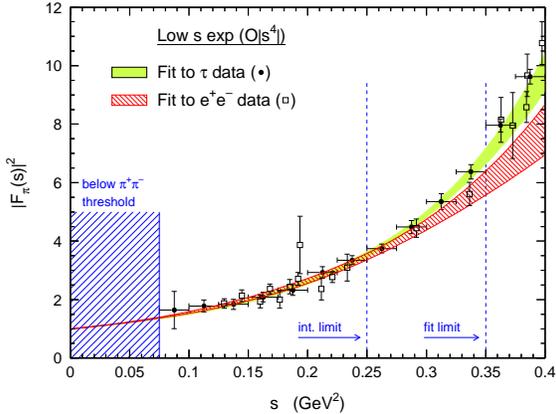}
\vspace{-0.7cm}
\caption[.]{Fit of the pion form factor from $4 m_\pi^2$ to
	$0.35~{\rm GeV}^2$ using a third-order Taylor expansion with 
	the constraints at $s=0$ and the measured
    	pion r.m.s. charge radius from space-like data~\cite{space_like}.
	The result of the fit is integrated only up to $0.25~{\rm GeV}^2$.}
\label{taylor}
\end{figure}

To overcome the lack of precise data at threshold energies and to 
benefit from the analyticity property of the pion form factor, 
a third order expansion in $s$ is used. The pion form factor $F_\pi^0$ 
is connected with the $\pi^+\pi^-$ cross section \via\ 
$|F^0_\pi|^2 = (3s/\pi\alpha^2 \beta_0^3)\,\sigma_{\pi\pi}$.
The expansion for small $s$ reads
\beq\label{eq_taylor}
F^0_{\pi} \;=\; 
      1 + \frac{1}{6}\langle r^2 \rangle_\pi\,s + c_1\,s^2 +c_2\,s^3 +
      O(s^4)~,
\eeq
where we use
$\langle r^2 \rangle_\pi=(0.439\pm0.008)~{\rm fm}^2$~\cite{space_like}, 
and the 
two parameters $c_{1,2}$ are fitted to the data in the range 
[$2m_\pi$, 0.6~GeV]. 
The results of the fits are explicitly quoted in Ref.~\cite{dehz}. We
show the functions obtained in Fig.~\ref{taylor}. Good agreement is 
observed  in the low energy region where the expansion should be 
reliable. Since the fits incorporate unquestionable constraints 
from first principles, we have chosen to use this parameterization 
for evaluating the integrals in the range up to 0.5~GeV. 

%
% ---------------
%
\subsection{Integration over the $\omega$ and $\phi$ Resonances}
\label{sec_omphi_int}

In the regions around the $\omega$ and $\phi$ resonances
we have assumed in the preceding 
works that the cross section of the $\pi^+\pi^-\pi^0$ production on the
one hand, and the $\pi^+\pi^-\pi^0$, $K^+K^-$ as well as $K^0_SK^0_L$
production on the other hand is saturated by the corresponding 
resonance production. In a data driven approach it is however more careful
to directly integrate the measurement points without introducing 
prior assumptions on the underlying process dynamics~\cite{teubner}. 
Possible non-resonant contributions and interference effects are thus 
accounted for. 

Notwithstanding,
a straightforward trapezoidal integration buries the danger of a bias:
with insufficient scan density, the linear interpolation of the 
measurements leads to a significant overestimation of the integral
when dealing with strongly concave functions such as the tails of 
Breit-Wigner resonance curves. 

We therefore perform a phenomenological fit of a BW resonance plus 
two Gaussians (only one Gaussian is necessary for the $\omega$)
to account for contributions other than the single
resonance. Both fits result in satisfactory $\chi^2$ values.
We have accounted for the systematics due to the arbitrariness 
in the choice of the parametrization by varying the functions 
and parameters used. The 
resulting effects are numerically small compared to the experimental
errors.

Since the experiments quote the 
cross section results without correcting for leptonic and hadronic 
vacuum polarization in the photon propagator, we perform the correction
here. 
The correction of hadronic vacuum polarization being iterative and 
thus only approximative, we assign half of the total vacuum polarization 
correction as generous systematic errors (\cf\   Section~\ref{sec_rad}). 
In spite of that, the evaluation of \amuhadLO\  is dominated by the 
experimental uncertainties. Since the trapezoidal rule is biased, we 
choose the results based on the BW fits for the final 
evaluation of \amuhadLO.

%
% ---------------
%
\subsection{Narrow $c\overline{c}$ and $b\overline{b}$ Resonances}
\label{sec_psi}

The contributions from the narrow $J/\psi$ resonances are computed
using a relativistic Breit-Wigner parametrization for their line shape.
The physical values for the resonance parameters and their errors are
taken from the latest compilation in Ref.~\cite{pdg2002}. Vacuum 
polarization effects are already included in the quoted leptonic widths. 
The total parametrization errors are then calculated by Gaussian error 
propagation. This integration procedure is not followed for the $\psi(3S)$
state which is already included in the $R$ measurements, and for the 
$\Upsilon$ resonances which are represented in an average sense (global 
quark-hadron duality) by the $b \overline{b}$ QCD contribution, discussed 
next.

%
% ---------------
%
\subsection{QCD Prediction at High Energy}
\label{sec_qcd}

Since the emphasis in this paper is on a complete and critical evaluation
of \sfs\ from low-energy data, we have adopted the conservative choice
of using the QCD prediction only above an energy of 5~GeV. The details of the 
calculation can be found in our earlier publications~\cite{dh97,dh98,dehz} 
and in the references therein. 

A test of the QCD prediction can be performed in the energy range between
1.8 and 3.7 GeV. The contribution to \amuhadLO\ in this region is computed
to be $(33.87\pm0.46)~10^{-10}$ using QCD, to be compared with the 
result, $(34.9\pm1.8)~10^{-10}$ from the data. The two values agree
within the $5\%$ accuracy of the measurements.

%
% ----------------------------- Results ------------------------------
%
\section{RESULTS}
\label{sec_results}

%
% ------------
%
\subsection{Lowest Order Hadronic Contributions}

We use the trapezoidal rule to integrate the experimental data
points (with the exception of the narrow resonances). 
Correlations between the measurements as well as among
experiments have been taken into account~\cite{dehz}.
Before adding up all the contributions to \amuhadLO, we shall summarize
the procedure. On the one hand, the \ee-based evaluation is done in three
pieces: the sum of exclusive channels below 2~GeV, the $R$ measurements
in the 2-5~GeV range and the QCD prediction for $R$ above. Major
contributions stem from the $2\pi$ (73\%) and the two $4\pi$ ($4.5\%$) 
channels. On the other hand, in the $\tau$-based evaluation, the latter 
three contributions are taken from $\tau$ data up to 1.6~GeV and 
complemented by \ee\ data above, because the $\tau$ \sfs\ run out 
of precision near the kinematic limit of the $\tau$ mass. Thus, 
for nearly $77\%$ of \amuhadLO\  (contributing $80\%$ of the 
total error-squared), two independent evaluations (\ee\ and $\tau$) 
are produced, the remainder being computed from \ee\ data and QCD alone.

\begin{figure*}[p]
\begin{center}
\includegraphics[width=34pc]{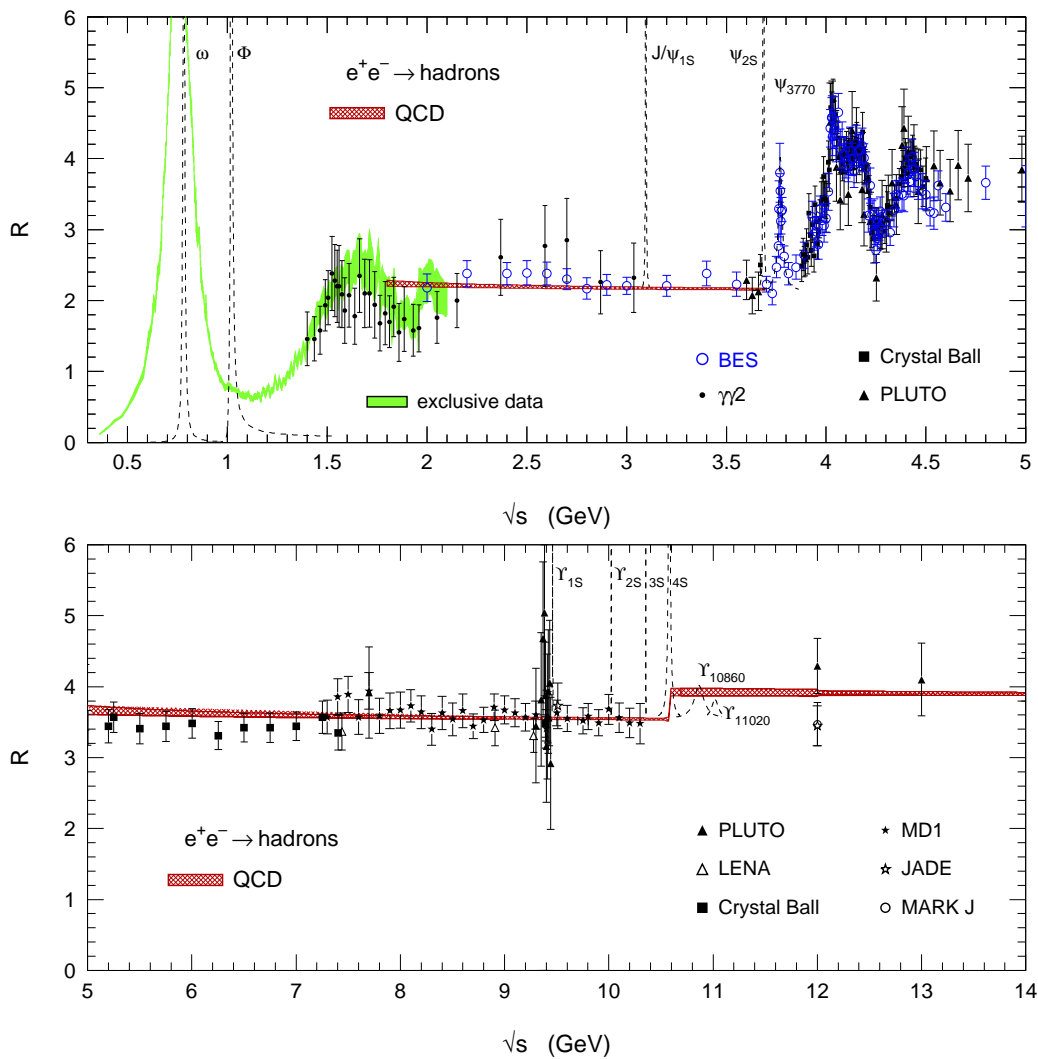}
\end{center}
\vspace{-0.7cm}
\caption[.]{Compilation of the data contributing to \amuhadLO. 
        Shown is the total hadronic over muonic cross section ratio $R$.
        The shaded band below 2~GeV represents the sum of the exclusive 
        channels considered in this analysis, with the exception of 
        the contributions from the narrow resonances which are given
        as dashed lines.
        All data points shown correspond to inclusive measurements. 
	The cross-hatched band gives the prediction
        from (essentially) perturbative QCD, which is found to be in good
        agreement with the measurements in the continuum 
	above 2~GeV. In this figure
        the $b\overline{b}$ threshold is indicated at the onset of 
        $B\overline{B}$ states in order to facilitate the comparison with 
        data in the
        continuum. In the actual calculation the threshold is taken at
        twice the pole mass of the $b$ quark.}
\label{fig_ree_all}
\end{figure*}
\begin{table*}[p]
\caption{Summary of the \amuhadLO\ contributions from \ee\ 
        annihilation and $\tau$ decays. The uncertainties 
	on the vacuum polarization
	and FSR corrections are given as second errors in the individual 
        \ee\ contributions, while those from isospin breaking are 
        similarly given for the $\tau$ contributions. These 'theoretical'
        uncertainties are correlated among all channels, except in the
        case of isospin breaking which shows little correlation between 
        the $2\pi$ and $4\pi$ channels. The errors given 
        for the sums in the last line are from the experiment, the missing 
        radiative corrections in \ee\ and, in addition for $\tau$, SU(2)
        breaking.}
\label{tab_results}
\setlength{\tabcolsep}{0.49pc}
{\normalsize
\begin{tabular}{lcrrr} \hline 
&&&& \\[-0.2cm]
 & & \mc{3}{c}{\amuhadLO\ ($10^{-10}$)} \\
\rs{Modes} & \mc{1}{c}{\rs{Energy [GeV]}} & \mc{1}{c}{~\ee} 
	& \mc{1}{c}{$~\tau$\,$^(\footnotemark[3]{^)}$} 
		& \mc{1}{c}{$~\Delta(e^+e^--\tau)$} \\[0.15cm]
\hline
&&&& \\[-0.3cm]
Low $s$ exp. $\pi^+\pi^-$
	& $[2m_{\pi^\pm}-0.500]$   & $ 58.04\pm1.70\pm1.14$  
			& $ 56.03\pm1.61\pm0.28$ & $ +2.0\pm2.6$ \\
$\pi^+\pi^-      $  
	& $[0.500-1.800]$    & $440.81\pm4.65\pm1.54$  
			& $464.03\pm3.19\pm2.34$
                        & $-23.2\pm6.3$ \\
$\pi^0\gamma$, $\eta \gamma$\,$^(\footnotemark[1]{^)}$
	& $[0.500-1.800]$    & $  0.93\pm0.15\pm0.01$  & -  & - \\
$\omega$          
	& $[0.300-0.810]$    & $ 36.94\pm0.84\pm0.80$  & -  & - \\
$\pi^+\pi^-\pi^0$  {\footnotesize[below $\phi$]}
	& $[0.810-1.000]$    & $  4.20\pm0.40\pm0.05$  & -  & - \\
$\phi$  
	& $[1.000-1.055]$    & $ 34.80\pm0.92\pm0.64$  & -  & - \\
$\pi^+\pi^-\pi^0$  {\footnotesize[above $\phi$]}
	& $[1.055-1.800]$    & $  2.45\pm0.26\pm0.03$  & -  & - \\
$\pi^+\pi^-2\pi^0       $  
	& $[1.020-1.800]$    & $ 16.73\pm1.32\pm0.20$  
		& $ 21.44\pm1.33\pm0.60$
			& $ -4.7\pm1.8$ \\
$2\pi^+2\pi^-           $  
	& $[0.800-1.800]$    & $ 13.95\pm0.90\pm0.23$  
		& $ 12.34\pm0.96\pm0.40$
			& $ +1.6\pm2.0$ \\
$2\pi^+2\pi^-\pi^0        $  
	& $[1.019-1.800]$    & $  2.09\pm0.43\pm0.04$  & -  & - \\
$\pi^+\pi^-3\pi^0 $\,$^(\footnotemark[2]{^)}$  
	& $[1.019-1.800]$    & $  1.29\pm0.22\pm0.02$  & -  & - \\
$3\pi^+3\pi^-    $  
	& $[1.350-1.800]$    & $  0.10\pm0.10\pm0.00$  & -  & - \\
$2\pi^+2\pi^-2\pi^0       $  
	& $[1.350-1.800]$    & $  1.41\pm0.30\pm0.03$  & -  & - \\
$\pi^+\pi^-4\pi^0       $\,$^(\footnotemark[2]{^)}$    
	& $[1.350-1.800]$    & $  0.06\pm0.06\pm0.00$  & -  & - \\
$\eta${\footnotesize($ \rar\pi^+\pi^-\gamma$, $2\gamma$)}$\pi^+\pi^-$ 
	& $[1.075-1.800]$    & $  0.54\pm0.07\pm0.01$  & -  & - \\
$\omega${\footnotesize($\rar\pi^0\gamma$)}$\pi^{0}$
	& $[0.975-1.800]$    & $  0.63\pm0.10\pm0.01$  & -  & - \\
$\omega${\footnotesize($\rar\pi^0\gamma$)}$(\pi\pi)^{0}$
	& $[1.340-1.800]$    & $  0.08\pm0.01\pm0.00$  & -  & - \\
$K^+K^-            $  
	& $[1.055-1.800]$    & $  4.63\pm0.40\pm0.06$  & -  & - \\
$K^0_S K^0_L         $  
	& $[1.097-1.800]$    & $  0.94\pm0.10\pm0.01$  & -  & - \\
$K^0K^\pm\pi^\mp         $\,$^(\footnotemark[2]{^)}$    
	& $[1.340-1.800]$    & $  1.84\pm0.24\pm0.02$  & -  & - \\
$K\overline K\pi^0$\,$^(\footnotemark[2]{^)}$    
	& $[1.440-1.800]$    & $  0.60\pm0.20\pm0.01$  & -  & - \\
$K\overline K\pi\pi         $\,$^(\footnotemark[2]{^)}$    
	& $[1.441-1.800]$    & $  2.22\pm1.02\pm0.03$  & -  & - \\
$R=\sum{\rm excl.~modes}    $  
	& $[1.800-2.000]$    & $  8.20\pm0.66\pm0.10$  & -  & - \\
$R$ {\footnotesize[Data]}
	& $[2.000-3.700]$    & $ 26.70\pm1.70\pm0.00$  & -  & - \\
$J/\psi$         
	& $[3.088-3.106]$    & $  5.94\pm0.35\pm0.03$  & -  & - \\
$\psi(2S)$ 
	& $[3.658-3.714]$    & $  1.50\pm0.14\pm0.00$  & -  & - \\
$R$ {\footnotesize[Data]}  
	& $[3.700-5.000]$    & $  7.22\pm0.28\pm0.00$  & -  & - \\
$R_{udsc}$ {\footnotesize[QCD]}
	& $[5.000-9.300]$    & $  6.87\pm0.10\pm0.00$  & -  & - \\
$R_{udscb}$ {\footnotesize[QCD]}
	& $[9.300-12.00]$    & $  1.21\pm0.05\pm0.00$  & -  & - \\
$R_{udscbt}$  {\footnotesize[QCD]}
	& $[12.0-\infty]$    & $  1.80\pm0.01\pm0.00$  & -  & - \\[0.15cm]
\hline
&&&& \\[-0.3cm]
	&
			     & \mc{1}{r}{$684.7\pm6.0_{\rm exp}~~~$}  
			     & \mc{1}{l}{$709.0\pm5.1_{\rm exp}$} 
			     & \\
\rs{$\sum\;(e^+e^-\rightarrow\:$hadrons)}
 	& \rs{$[2m_{\pi^\pm}-\infty]$}
			     & \mc{1}{r}{$\pm\,3.6_{\rm rad\,}~~~$}  
			     & \mc{1}{r}{$\pm\,1.2_{\rm rad}\pm2.8_{\rm SU(2)}$} 
			     & \mc{1}{r}{\rs{$-24.3\pm7.9_{\rm tot}$}} 
	
\\[0.15cm]
 \hline
\end{tabular}
}
\vspace{0.0cm}
{\footnotesize 
\begin{quote}
$^{1}\,$Not including $\omega$ and $\phi$ resonances (see text). \\ \noindent
$^{2}\,$Using isospin relations (see text). \\ \noindent
$^{3}\,$\ee\  data are used above 1.6~GeV (see text). \\ \noindent
\end{quote}
} 
\end{table*}
Fig.~\ref{fig_ree_all} gives a panoramic view of the \ee\ data in the 
relevant energy range. The shaded band below 2~GeV represents the sum 
of the exclusive channels considered in the analysis. It turns out to be
smaller than our previous estimate~\cite{adh}, 
essentially because more complete data sets are used and new information 
on the dynamics could be incorporated in the isospin
constraints for the missing channels. The QCD prediction is indicated by the 
cross-hatched band. It is used in this analysis only for energies 
above 5~GeV. Note that the QCD band is plotted taking into account 
the thresholds for open flavour $B$ states, in order to facilitate 
the comparison with the data in the continuum. However, for the 
evaluation of the integral, the $b\overline{b}$ threshold is taken 
at twice the pole mass of the $b$ quark, so that the contribution 
includes the narrow $\Upsilon$ resonances, according to global 
quark-hadron duality.

The contributions from the different processes in their indicated 
energy ranges are listed in Table~\ref{tab_results}.
Wherever relevant, the two \ee- and $\tau$-based evaluations are given.
The discrepancies discussed above
are now expressed directly in terms of \amuhadLO\, giving smaller
estimates for \ee\ data by 
$(-21.2\pm6.4_{\rm exp}\pm2.4_{\rm rad}\pm2.6_{\rm SU(2)}\,(\pm7.3_{\rm total}))~10^{-10}$ for the $2\pi$ channel and 
$(-3.1\pm2.6_{\rm exp}\pm0.3_{\rm rad}\pm1.0_{\rm SU(2)}\,(\pm2.9_{\rm total}))~10^{-10}$ for the sum of the $4\pi$ channels. 
The total discrepancy  
$(-24.3\pm6.9_{\rm exp}\pm2.7_{\rm rad}\pm2.8_{\rm SU(2)}\,(\pm7.9_{\rm total}))~10^{-10}$ amounts to 3.1 standard deviations and
precludes from performing a straightforward combination of the two 
evaluations.

%
% -------------
%
\subsection{Results for $a_\mu$}
\label{sec_results_amu}
\begin{figure}[t]
\hspace{-1.0cm}
\includegraphics[width=20.0pc]{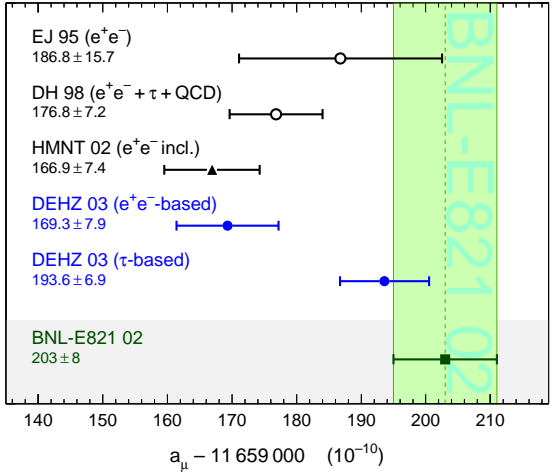}
\vspace{-0.7cm}
\caption[.]{Comparison of the results with the 
	BNL measurement~\cite{bnl_2002}. Also shown
	are our previous estimates~\cite{eidelman,dh98} obtained before 
	the CMD-2 data were available, and the recent evaluation of 
	Hagiwara {\em et al.}~\cite{teubner}.}
\label{fig:results}
\end{figure}

The results for the lowest order hadronic contribution are 
\beqns
  a_\mu^{\rm had,LO}\!\!\! &=&\!\!\! 
	(684.7\pm6.0_{\rm exp}\pm3.6_{\rm rad})~10^{-10}~,\\
 a_\mu^{\rm had,LO}\!\!\! &=&\!\!\! 
	(709.0\pm5.1_{\rm exp}\pm3.0_{\rm rad,SU(2)})~10^{-10}~,
\eeqns
where the first numbers are \ee-based and the second $\tau$-based.
Adding the QED, higher-order hadronic, light-by-light scattering and
weak contributions as given in Section~\ref{anomaly},
we obtain for $a_\mu$
\beqns
  a_\mu^{\rm SM} &=& (11\,659\,169.3\pm7.0_{\rm had}\pm3.5_{\rm LBL} \\
	&&\hspace{2.0cm}	\pm\;0.4_{\rm QED+EW})~10^{-10}~, \\
 a_\mu^{\rm SM} &=& 	(11\,659\,193.6\pm5.9_{\rm had}\pm3.5_{\rm LBL}\\
	&&\hspace{2.0cm}	\pm\;0.4_{\rm QED+EW})~10^{-10}~,
\eeqns
where again the first results are \ee-based and the second $\tau$-based.
These values can be compared to the present experimental average given 
in Eq.~(\ref{bnl}). Adding experimental and theoretical errors 
in quadrature, the differences between measured and computed values 
are found to be (first number \ee-based and the second $\tau$-based)
\beqns
  a_\mu^{\rm exp}-a_\mu^{\rm SM} \,&=&\, (33.7\pm11.2)~10^{-10}~, \\
  a_\mu^{\rm exp}-a_\mu^{\rm SM} \,&=&\, (9.4\pm10.5)~10^{-10}~,
\eeqns
corresponding to 3.0 and 0.9 standard deviations, respectively.
A graphical comparison of the results with the 
experimental value is given in Fig.~\ref{fig:results}. Also shown
are our previous estimates~\cite{eidelman,dh98} obtained before 
the CMD-2 and the new $\tau$ data were available (see discussion below), 
and the recent evaluation of Hagiwara {\em et al.}~\cite{teubner}.

%
% ----------------------------- Discussion ------------------------------
%
\section{DISCUSSION}
\label{sec_discuss}

%
% -------------
%
\subsection{The Problem of the $2\pi$ Contribution}
\label{sec_discuss_2pi}

The significant discrepancy between the \ee\ and $\tau$ evaluations of 
\amuhadLO\ is a matter of concern. The following changes in the dominant
$\pi^+\pi^-$ contribution (all expressed 
in $10^{-10}$ units) are observed with respect to our earlier work~\cite{dh98}:
\bei

\item 	the new CMD-2 data~\cite{cmd2} produce a downward shift of the 
	\ee\ evaluation by 1.9 (well within errors from previous experiments),

\item 	the new ALEPH data~\cite{aleph_new} increases the $\tau$ evaluation 
	by 3.5,

\item 	including the CLEO data in the $\tau$ evaluation slightly improves the 
	precision, but further raises the central value by 4.0,

\item 	although including the OPAL data has little effect on the overall
	precision, it also increases the result by 1.9,

\item 	the new complete isospin symmetry-breaking correction,
	including the re-evaluation of the $S_{\rm EW}$ factor,
	increases the $\tau$ evaluation by $0.2$.

\eei
In principle, the observed discrepancy for the $2\pi$ contribution,
$(-21.2\pm7.3)$, or $(-4.2\pm1.4)$\% when expressed 
with respect to \ee, could be caused by any (or the combination of 
several) of the following three effects which we examine in turn:
\bei

\item {\bf The normalization of \ee\  data}\\[0.05cm]
Here, as below, 'normalization' does not necessarily mean an overall
factor, but refers to the absolute scale of the 'bare' cross section
at each energy point. There is no cross check of this at the precision
of the new CMD-2 analysis. The only test we can provide is to compute
the \ee\ integral using the experiments separately. Because of the 
limited energy range where the major experiments overlap, we choose
to perform the integration in the range of $\sqrt{s}$ from 610.5 to 820~MeV.
The corresponding contributions are:
$313.5\pm3.1$ for CMD-2, $321.8\pm13.9$ for OLYA, $320.8\pm12.6$ for CMD,
and $323.9\pm2.1$ for the isospin-corrected $\tau$ data. 
No errors on radiative corrections and isospin breaking are
included in the above results.

\item {\bf The normalization of $\tau$ data}\\[0.05cm]
The situation is quite similar, as the evaluation is dominated by the
ALEPH data. It is also possible to compare the results provided by each
experiment separately, with the \sfs\ normalized to the respective
hadronic branching ratios. Leaving aside the region below 500~MeV where a fit
combining analyticity constraints is used, the contributions are:
$460.1\pm4.4$ for ALEPH, $464.7\pm9.3$ for CLEO and 
$464.2\pm8.1$ for OPAL, where the common error 
on isospin breaking has been left out.
The three values are consistent with each other and even the less 
precise values are not in good
agreement with the \ee\ estimate in this range, $440.8\pm4.7$,
not including the error on missing radiative corrections.

Apart from an overall normalization effect, differences could originate
from the shape of the measured \sfs. If all three \sfs\ are normalized to
the world average branching ratio (our final procedure), then the results
for the contribution above 0.5~GeV become:
$459.9\pm3.6$ for ALEPH, $465.4\pm5.1$ for CLEO and 
$464.5\pm5.1$ for OPAL, with a common error of $\pm$2.4 from the $\pi \pi^0$ 
and leptonic branching ratios and the uncertainty on isospin breaking 
left out. Again the results are consistent and their respective experimental 
errors give a better feeling of the relative impact of the measurements.

\item {\bf The isospin-breaking correction applied 
	   to $\tau$ data}\\[0.05cm]
The basic components entering SU(2) breaking have been identified. The
weak points before were the poor knowledge of the long-distance 
radiative corrections and the quantitative effect of loops. 
Both points have been addressed by the analysis of
Ref.~\cite{ecker2} showing that the effects are small and covered by
the errors previously applied. The overall effect of the 
isospin-breaking corrections (including FSR)
applied to the $2\pi$ $\tau$ data, expressed in relative terms, is
$(-1.8\pm0.5)\%$. Its largest contribution ($-2.3\%$) 
stems from the uncontroversial
short-distance electroweak correction. Additional contributions 
must be identified to bridge the observed difference.

\eei

Thus we are unable at this point to identify the source of the discrepancy.
More experimental and theoretical work is needed. On the experimental side,
additional data is available from CMD-2, 
but not yet published. As an alternative, a promising approach 
using \ee\  annihilation events with initial state radiation (ISR), 
as proposed in Ref.~\cite{kuehn_isr}, allows a single experiment 
to cover simultaneously
a broad energy range. Two experimental programs are underway at Frascati
with KLOE~\cite{kloe_isr} and at SLAC with BABAR~\cite{babar_isr}. The
expected statistics are abundant, but it will be a challenge to reduce
the systematic uncertainty at the level necessary to probe the CMD-2
results. As for $\tau$'s,
the attention is now focused on the forthcoming results from the $B$ 
factories. Again, the quality of the analysis will be determined 
by the capability to control systematics rather than the already sufficient
statistical accuracy.
On the theory side, the computation of more precise and more
complete radiative corrections both for \ee\ cross sections and 
$\tau$ decays should be actively pursued.  

Other points of discussions
are pursued in Ref.~\cite{dehz}.

%
% -------------
%
\subsection{Consequences for \aqedZ}
\label{sec_results_alpha}

In spite of the fact that the present analysis was focused on the
theoretical prediction for the muon magnetic anomaly, it is possible
to draw some conclusions relevant to the evaluation of the hadronic vacuum
polarization correction to the fine structure constant at $M_Z^2$. 
The problem found in the $2\pi$ \sf\  is less important for 
\daqedZ\  with respect to the total uncertainty,
because the integral involved gives less weight to the low-energy region. 
The difference between the evaluations using the $2\pi$, 
$4\pi$ and $2\pi2\pi^0$ \sfs\ from \ee\ and $\tau$ data are found to be:
\beqns
 \Delta\alpha^{ee}_{\rm had}(M_{Z}^2)-
 \Delta\alpha^{\tau}_{\rm had}(M_{Z}^2)= (-2.8\pm0.8)~10^{-4}~.
\eeqns
While this low-energy contribution shows a 3.5 standard deviation 
discrepancy (when adding the different errors in quadrature), it also 
exceeds the total uncertainty of $1.6~10^{-4}$ 
on \daqedZ\ which was quoted in Ref.~\cite{dh98}. It is worth pointing out
that such a shift produces a noticeable effect for the determination of the 
Higgs boson mass $M_{\rm H}$ in the global electroweak fit~\cite{haidt}. 
With the present input for the electroweak observables~\cite{lepewwg} 
from LEP, SLC and FNAL yielding central values for $M_{\rm H}$ around 100~GeV,
going from the \ee\- to the $\tau$-based evaluation induces
a decrease of $M_{\rm H}$ by 16~GeV using all observables and by 20~GeV when
only the most sensitive observable, $(\sin^2\theta_{\rm W})_{\rm eff}$, 
is used.
  
%
% --------------------------- Conclusions -----------------------------
%
\section{CONCLUSIONS}

A new analysis of the lowest-order hadronic vacuum polarization contribution
to the muon anomalous magnetic moment has been presented. It is based on the
most recent high-precision experimental data from \ee\ annihilation and 
$\tau$ decays in the $\pi\pi$ channel. Special attention was given to 
the problem of isospin symmetry breaking and the corresponding 
corrections to be applied to $\tau$ data. 
The main results of our analysis are the following:

\bei
\item 	the new evaluation based solely on \ee\ data is significantly
	lower than previous estimates and is in conflict with the experimental
	determination of $a_\mu$ by 3.0 standard deviations.

\item 	the new precise evaluations of the dominant $\pi\pi$ contributions 
	from \ee\ annihilation and isospin-breaking corrected $\tau$ decays are 
	not anymore in agreement with each other. A discussion has been presented 
	for possible sources of the discrepancy which could not be resolved.
	This situation is a matter of great concern, as the $\tau$-based 
	prediction of $a_\mu$ is in better agreement with the experimental 
	value, from which it deviates by non-significant 0.9 standard 
	deviations.
\eei
More experimental and theoretical work is needed to lift the present
uncertainty on whether or not new physics has been uncovered with the
muon magnetic moment.  
% 
% --------------------------- Acknowledgement -------------------------
%
\subsection*{Acknowledgements}

I am indepted to my collegues M.~Davier, S.~Eidelman and Z.~Zhang 
for the pleasent collaboration.
The close collaboration with V.~Cirigliano, G.~Ecker and H.~Neufeld 
is greatfully acknowledged. Discussions with F.~Jegerlehner, 
J.~H.~K\"uhn, A.~Pich, A.~Stahl, A.~Vainshtein and especially 
W.~Marciano are appreciated. We thank S.~Menke for providing us 
with the OPAL spectral functions.
Finally, many thanks to Abe Seiden and his team for the preparation of 
the Tau'02 workshop.

%
% ------------------------- Bibliography ------------------------------
%
\end{document}